\documentclass[referee,epsf]{aa} 

\newcounter{sub}
\newcounter{subeqn}[sub]
\newcommand\be{\begin{equation}}
\newcommand\ee{\end{equation}}
\newcommand\lp{\left(}
\newcommand\rp{\right)}
\newcommand\ls{\left[}

\newcommand\rs{\right]}

\newcommand\st{\stepcounter{sub}}
\newcommand\stq{\stepcounter{subeqn} }
\newcommand\bea{\begin{eqnarray}}
\newcommand\eea{\end{eqnarray}}
\newcommand\bean{\begin{eqnarray*}}
\newcommand\eean{\end{eqnarray*}}

\newcommand\oomega{\mbox{\boldmath $\omega$}}

\newcommand\unit{\mbox{\boldmath $e$}}

\newcommand\m{{\bf m}}

\newcommand\I{{\bf I}}
\newcommand\LL{{\bf L}}

\newcommand\T{{\bf T}}
\newcommand\no{\nonumber}

\begin{document}

   \thesaurus{06     
              (
                19.63.1)} 
   \title{On the precession of the isolated pulsar PSR B1828-11}

   \subtitle{A time-varying magnetic field}

   \author{Vahid Rezania }

   \offprints{V. Rezania \\ email: vrezania@phys.ualberta.ca\\
                   $^*$ Present address}

    \institute{Department of Physics, University of Alberta,
              Edmonton AB, Canada T6G 2J1 $^*$ \\       
              Institute for Advanced Studies in Basic Sciences,
               Gava Zang, Zanjan 45195, Iran\\
                    }

   \date{Received / accepted }

\maketitle

\abstract{
Analysis of both pulse timing and pulse shape variations of the isolated 
pulsar PSR B1828-11 shows highly correlated and strong Fourier power at 
periods $\simeq 1000, 500, 250$, and $167$ d (Stairs et al. \cite{SLS00}).
The only
description based on a free precession of star's rigid crust coupled 
to the magnetic dipole torque, explains the $500$-component, as the 
fundamental Fourier frequency, with its harmonic $250$-component 
(Link \& Epstein \cite{LE01}).  In this paper, we study a time-varying
magnetic field model and show that {\it if}
the dipole moment
vector rotates with a period {\it nearly equal} to the longest 
(assumed fundamental) observed period ($\simeq 1000$ d) relative to
the star's body axes, the resulting magnetic torque
may produce the whole Fourier spectrum consistently.  
We also find the second and fourth harmonics at periods $\simeq 500$ and 
$250$ d are dominant for small wobble angle 
$\simeq 3^\circ$ and large field's inclination 
angle $\geq 89^\circ$.
}

\section{Introduction}
The monitoring of long-term and periodic variations both in pulse shape and 
slow-down rate of the isolated pulsar PSR B1828-11 shows strong Fourier power
at periods of~$\simeq 1000, 500, 250$, and $167$ d, with the strongest
one at period~$\simeq 500$ d (Stairs et al. \cite{SLS00}).
The close relationship between
the periodic changes in the beam shape and the spin-down rate of the pulsar 
suggests the possibility of precession of the spin axis in a rotating
body.
The precession of the spin axis 
would provide cyclic changes in the inclination angle $\chi$ between the 
spin and magnetic symmetry axes.  The result will be periodic variations
both
in the observed pulse-profile and spin-down rate of the pulsar.   

Recently Jones \& Andersson (\cite{JA01}) and Link \& Epstein (\cite{LE01}) 
studied a freely precessing neutron 
star due to a small deformation of the star from spherical symmetry coupled 
to a torque 
such as magnetic dipole moment, gravitational radiation, etc., and 
explained some part of the observed data.  Because of the strong 
periodicity at period~$\simeq 504$ d seen in the data,
Jones \& Andersson (\cite{JA01})
reasonably suggested that the actual free precession period is $P_{\rm 
pre}=1009$ d.  A coupling between the magnetic dipole moment and star's 
spin axis can provide a strong modulation at period $P_{\rm pre}/2\simeq 
504$ d, when the magnetic dipole is nearly orthogonal to the star's 
deformation axis.  But their model could not explain the strong Fourier 
component 
corresponding to a period of $\simeq 250$ d
(see Stairs et al. \cite{SLS00}).
The latter component has a
significant contribution in the observed variations of period residual 
$\Delta p$, its derivative $\Delta \dot p$, and pulse shape.  For this 
reason, Link \& Epstein (\cite{LE01}) assumed that the strongest Fourier 
component ($\simeq 500$ d) represent the actual free precession period.
They found that
for a small deformation parameter of $\epsilon=(I_3-I_1)/I_1\simeq 9\times 
10^{-9}$, a free precession of the angular momentum axis around the 
symmetry axis of the 
crust could provide a period at $P_{\rm pre}\simeq 511$ d.  Here 
$I_1=I_2<I_3$ are the principle moment inertia of the star.   Further, 
they showed that a coupling of nearly orthogonal (fixed to
the body of the star) magnetic dipole
moment to the 
spin axis would provide the observed harmonic at period $\simeq 250$ 
d. Their model has good agreement with observations in the 
pulse period, but as they mentioned, it failed to 
explain the Fourier component at period $\simeq 1000$ d seen in the
data (as well as $167$ d).

The existence of precession in a neutron star is in strong conflict with 
the superfluid models for the neutron star interior structure.  These 
models have successfully explained the glitch phenomena (with both
pre- and post-glitch behavior) in most neutron stars in which the
pinned vortices
to the star crust become partially unpinned during a glitch
(Alpar et al. \cite{Alp84}).
As shown by Shaham (\cite{Sha77}) and Sedrakian et al. (\cite{Sed99}),
the precession
should be
damped out by the pinned (even imperfect) vortices on a time scale of few 
precession periods.   For example, PSR B1828-11 with typical degree of 
vortex pinning, $I_{\rm pinned}/I_{\rm star}\sim 1.4$\% (indicated by 
pulsar glitches in stars that frequently glitch), would precess for $\ll 40$ 
sec, far shorter than the observed periods (Link \& Epstein \cite{LE01}).
Here $I_{\rm
star}$ is the total moment inertia of the star, while $I_{\rm pinned}$ is 
the portion of star's fluid moment inertia that is pinned to the crust.   

The free precession 
description provides an effective decoupling between the internal 
superfluid and the crust. 
Recently, Link \& Cutler (\cite{LC01}) studied the problem more
carefully by considering
dynamics of the pinned vortices in a free precessing star under both Magnus,
$f_m$, and hydrodynamics (due to the precession), $f_p$, forces.
They found that the precessional (free) motion itself prevents the vortex 
pinning process and keeps the vortices unpinned in the crust of PSR 
B1828-11 while precessing, for a force density (per unit length) $f_p\sim 
10^{16}$ dyn cm$^{-1}$.    As a result, they found that partially 
pinned-vortex configuration cannot be static. 

The effective core-crust decoupling causes the core and the crust to
rotate at different rates (Sedrakian et al. \cite{Sed99}).\footnote{Actually 
Sedrakian et al. (\cite{Sed99})
have shown that even for partially pinned vortices the core and the
crust would rotate at different angular velocities.}  The latter
would increase core magnetic flux-tube displacement relative to the
crust, and then sustain the magnetic
stresses on the solid crust in forcing it to break (in platelets) and
move as the star rotate
(Ruderman \cite{Rud91ab}).
The stresses are strong enough to move the crustal lattice by continual
cracking, buckling, or plastic flow to relative stresses beyond the lattice
yield strength.  As a result, because of the very high electrical
conductivity of the crust ($\sigma \sim 10^{27}$ s$^{-1}$), the foot points
of external magnetic field lines move
with the conducting plates in which the field is entangled.   Furthermore,
since the core magnetic flux tubes are frozen into the core's fluid, the
precessing crust drags them and then increases core flux-tube displacements.
Therefore, during precession of the crust such plate motion is unavoidable.
In addition,
as shown by Malkus (\cite{Mal63}, \cite{Mal68}) the precessional motion of
the star exerts
torques
to the core and/or crust resulting from shearing flow at the thin core-crust
boundary region.  These
torques, so-called precessional torque, are able to sustain a turbulent
hydromagnetic flow in the boundary region, and then increase 
local magnetic field strength.  This would excite convective
fluid motions in the core-crust boundary, increase magnetic
stresses on the crust and cause it to break down in platelets.

In this paper, motivated by the above conjecture, we suggest that the
magnetic field may vary somewhat with time,
relative to the body axes of the star.
The question that arises now is whether
the whole observed Fourier spectrum of PSR B1828-11,
can be consistently generated by a {\it time-varying} magnetic
field during the course of free precession of the star. 
In other words, under what conditions will the observed cyclical changes
in the timing data be produced by
precession of the star's crust coupled to the magnetic dipole torque of    
a time-varying magnetic field.  In section \ref{precess} we
address this question in detail.  Following Link \& Epstein (\cite{LE01})
we assume 
that the star precesses freely 
around the spin axis, but with period $P_{\rm pre}\simeq 1000$ d.   Then we 
show that the magnetic torque exerted by a dipole moment
may produce the other observed harmonics as seen in data, {\it if} the
magnetic dipole vector rotates with a period close to $P_{\rm pre}$
relative to the star's body axes.
Section \ref{discuss} is devoted to further discussion.


\section{The precession}\label{precess}

Consider a rigid, biaxial rotating star with 
the principle axes $\unit_1, \unit_2, \unit_3$ and corresponding principle 
moment of inertia $I_1=I_2\neq I_3$.  The star's angular momentum $\LL$ is 
misaligned to the symmetry axis $\unit_3$ by a wobble angle $\theta$, ie. 
$\LL\cdot \unit_3=L\cos\theta$.  In general, we assume that the stellar
magnetic field is
dipolar and changes with time as
\st\bea\label{m}
&&\m=m_0\sin\chi f_1(t)\unit_1+m_0\sin\chi f_2(t) 
\unit_2 + m_0 \cos\chi f_3(t) \unit_3,
\eea
where $m_0$ is the
average
value of $|\m|$ over a period $2\pi/\omega_p$ and $\chi$ is the angle
between $\m$ and $\unit_3$. 
The functions $f_1(t), f_2(t)$ and $f_3(t)$ are arbitrary functions in time
and will be determined later by 
using the data.
The equations of 
motion in the corotating frame are 
\st\be
\label{eom-1}
\I\cdot\frac{d\oomega}{dt}+\oomega\times\LL
=\frac{2\omega^2}{3c^3}(\oomega\times\m)\times\m - 
\frac{1}{5Rc^2}(\oomega\cdot\m)
(\oomega\times\m),
\ee
where $R$ is the average radius of the star.
The first term of the magnetic torque, $\T_{\rm ff}$, is due to the 
far-field radiation and has components both parallel and perpendicular
to the
spin axis.  It is responsible for spinning down the star. The second term, 
$\T_{\rm nf}$, represents the near-field radiation torque and is exactly
perpendicular to the spin axis. It has no contribution to the energy/angular
momentum transfer from the star.  This torque does affect the wobble angle
and spin 
rate of a freely 
precessing star.
Following Link \& Epstein (\cite{LE01}), for small wobble angle 
$\theta\simeq 3^\circ$,
suggested by
the observed pulse shape variations of PSR B1828-11 over one precession 
period, and for small
oblateness $\epsilon\simeq 10^{-8}$, we have $(\omega_0\tau_{\rm ff})^{-1}\ll 
(\omega_0\tau_{\rm nf})^{-1} < \epsilon\theta \ll \theta \ll 1$.  
Here $\omega_0$ is the angular frequency of the star,
$\tau_{\rm nf}$ ($\sim 10^4$ yr), and
$\tau_{\rm ff}$ ($\sim 10^8$ yr) are the corresponding near- and far-field
radiation torque time scales, respectively.
So up to the first 
order of $\theta$, the magnetic torques in the RHS of equation 
(\ref{eom-1}) can be neglected.   
Therefore in this order, the angular velocity vector
$\oomega$ precesses freely around the star's symmetry axis as 
\st\be\label{sol}
\oomega(t)\simeq \theta\omega_0\cos(\omega_p t+\beta_p)\unit_1
 + \theta\omega_0 \sin(\omega_p t+\beta_p)\unit_2
 + \omega_0 \unit_3,
\ee
with the precession frequency $\omega_{p}=\epsilon\omega_0$,
and a constant phase $\beta_p$.
For the case of PSR B1828-11, observations 
suggest that $\epsilon \simeq 4.7\times 10^{-9}$
(Stairs et al. \cite{SLS00}).
Then $\omega_{p}\simeq 7.29\times 10^{-8}$ Hz or equivalently 
$P_{\rm pre}=2\pi/\omega_{p}\simeq 997$ d. 


\subsection{Timing}
The observed timing behavior can be understood by considering the
contribution
of other torque's components in the variations of the spin rate.   
By multiplying
$\oomega$ in equations (\ref{eom-1}) we have 
\st\be\label{wT}
\frac{d\omega^2}{dt}=\frac{2}{I_1}\lp\frac{}{}\oomega\cdot\T -  
        \epsilon\frac{I_1}{I_3}\omega_3 T_3\rp.
\ee
Equation (\ref{wT}) shows the torque-induced variations in the spin rate of 
star. 
From equation (\ref{eom-1}) it is clear that the $\oomega\cdot\T$ term
does not
depend on the near-field torque.  It contributes to the spin rate change 
only through the negligible final term. 
Using equations (\ref{m}) and (\ref{eom-1}), by calculating 
$\oomega\cdot\T_{\rm ff}$ 
and subtracting equation (\ref{wT}) from the secular spin down of the 
star (in the absence of precession), 
$-(\omega\sin^2\chi/\tau_{\rm ff}) (I_3/I_1)$, 
one can find the spin rate due to the far-field torque as 
(dropping constant terms)
\st\bea\label{spin-var-mag1}
&&\frac{\Delta\dot\omega}{\omega_0} \simeq
\frac{1}{\tau_{\rm ff}}\frac{I_3}{I_1} 
\ls \frac{}{} \cos^2\chi f^2_3(t)
+~\theta\sin2\chi\lp \frac{}{} 
\cos(\omega_pt+\beta_p) f_1(t)f_3(t) + \sin(\omega_pt+\beta_p) f_2(t)f_3(t) 
\frac{}{}\rp
\right. \no\\ 
&&\hspace{3cm} 
- \frac{\theta^2}{2} \sin^2\chi \lp \frac{}{} 
\sin^2(\omega_pt+\beta_p) f^2_1(t) 
+\cos^2(\omega_pt+\beta_p) f_2^2(t)\right.\no\\
&&\hspace{5.5cm}\left.\left. 
- 2 \sin(\omega_pt+\beta_p)\cos(\omega_pt+\beta_p)f_1(t) f_2(t)
\frac{}{}\rp\frac{}{}\rs.
\eea
Using Fourier expansion, we expand the functions $f_1(t), f_2(t)$, and
$f_3(t)$ as follow 
\stq\bea\label{f}
&&f_1(t)=\sum_{n=0} a_n\cos(n\omega_d+n\beta_d),\no\\
&&f_2(t)=\sum_{n=0} b_n\sin(n\omega_d+n\beta_d),\no\\   
&&f_3(t)=\sum_{n=0} c_n\cos(n\omega_d+n\beta_d),
\eea
where $\omega_d$ is frequency of the magnetic field's variation and
$\beta_d$ is constant. The coefficients $a_n, b_n$, and $c_n$ will be
determined by fitting the data.
Equation (\ref{spin-var-mag1}) shows that the spin rate variations depend 
on both the precession frequency and the variation frequency of the dipole
field.
For the case $a_0=c_0=1$ and $a_n=b_{n-1}=c_n=0$ for $n\geq 1$, equation
(\ref{spin-var-mag1}) reduces to one obtained by
Link \& Epstein (\cite{LE01})
(except by a factor $I_3/I_1$). 

To obtain the reported spectrum of PSR B1828-11, it is enough to consider 
$n=0$ and $1$ terms in equation (\ref{spin-var-mag1}) only.  
The $n\geq 2$ terms will produce 
the higher harmonics which will be discussed later.  Simply a correct
behavior of the spin rate can be found by setting $c_1=0$, $c_0=1$, and
$b_1=-a_1$.  
Therefore equation (\ref{spin-var-mag1}) reduces to (dropping constant terms) 
\st\bea\label{spin-var-mag2}
&&\frac{\Delta\dot\omega}{\omega_0} \simeq
\frac{\theta}{\tau_{\rm ff}}\frac{I_3}{I_1} 
\ls \frac{}{} \sin2\chi\lp \frac{}{} 
a_0\cos(\omega_pt+\beta_p) +a_1 \cos[(\omega_p+\omega_d)t+\beta_p +\beta_d] 
\frac{}{}\rp
\right. \no\\ 
&&\hspace{2.5cm}\left. 
-\frac{\theta}{2} \sin^2\chi \lp \frac{}{} 
2a_0a_1\cos(\omega_dt+\beta_d) 
- 2a_0a_1\cos[(2\omega_p+\omega_d)t +2\beta_p+ \beta_d] 
\right. \right.\no\\ 
&&\hspace{3.5cm} 
-a_0^2\cos(2\omega_pt+2\beta_d)
\left. \left. 
- a_1^2\cos[2(\omega_p + \omega_d)t+  2(\beta_p+\beta_d)] \frac{}{}\rp
\frac{}{}\rs.
\eea

As one expected, the expression for observable variations in period 
derivative, $\Delta \dot p$, will be modified as well.  
The star's residual in $\dot p$ is owing to both 
torque variation, equation (\ref{spin-var-mag1}), and the geometrical 
effect.  
The later is due to the orientation of the star's angular velocity vector 
$\oomega$ relative to the observer.  As expected, the torque 
effects dominate the geometrical effects by a factor $(P_{\rm pre}^2 /\pi 
P_0 \tau_{\rm ff})(I_3/I_1)\sin^2\chi\simeq 100-1000$ for the precession 
period $\simeq 1000 $ d, and so we neglected it here.  Therefore
\st\bea\label{timing}
&&\Delta\dot{p}
\simeq - \frac{P_0^2}{2\pi}{\Delta\dot\omega}
\simeq -{P_0\over T}\theta
\ls \frac{}{} \sin2\chi\lp \frac{}{} 
a_0\cos(\omega_pt+\beta_p) +a_1 \cos[(\omega_p+\omega_d)t+\beta_p +\beta_d] 
\frac{}{}\rp
\right. \no\\ 
&&\hspace{2.5cm}\left. 
-\frac{\theta}{2} \sin^2\chi \lp \frac{}{} 
2a_0a_1\cos(\omega_dt+\beta_d) 
- 2a_0a_1\cos[(2\omega_p+\omega_d)t +2\beta_p+ \beta_d] 
\right. \right.\no\\ 
&&\hspace{3.5cm} 
-a_0^2\cos(2\omega_pt+2\beta_d)
\left. \left. 
- a_1^2\cos[2(\omega_p + \omega_d)t+  2(\beta_p+\beta_d)] \frac{}{}\rp
\frac{}{}\rs,
\eea
where $T=(\tau_{\rm ff}/2\sin^2\chi)(I_1/I_3)\simeq t_{age}$ is 
approximately equal to the characteristic spin-down age and $P_0$ is the 
spin period of star.  Equation (\ref{timing}) gives the period derivative 
residual due to far-field torque variations.  The observations showed that 
both $500$ and $250$ Fourier components 
are dominant and have comparable amplitudes 
(Stairs et al. \cite{SLS00}, Link \& Epstein \cite{LE01}).   
By assuming the precession frequency as $\omega_p\simeq 2\pi/500$ d$^{-1}$, 
one can get the Link \& Epstein's results for $|a_0|\gg |a_1|$ and 
$\omega_d\ll\omega_p$. But 
by setting $\omega_p+\omega_d\simeq 2\pi/508$ d$^{-1}$ (chosen by fitting 
the data), or equivalently $1/P_{\rm pre}+1/P_d\simeq 1/508$ d$^{-1}$,
we have
$P_d/P_{\rm pre}\simeq (P_{\rm pre}/508-1)^{-1}$ where $P_d$ and $P_{\rm 
pre}$ measured in days.  Unfortunately the $1000$ Fourier component in the 
data is not as accurate as the other components.  So for $P_{\rm 
pre}\simeq 1016$ d, we have $P_d\simeq P_{\rm pre}$ or $\omega_d\simeq 
\omega_p$.  The value of $P_d$ will be larger (lower) than $P_{\rm pre}$, if 
$P_{\rm pre}$ ($>508$ d)  is lower (larger) than $1016$ d.
Therefore equation (\ref{timing}) 
reduces to (for $\beta_p=0$ and  $\beta_d=0$)
\st\bea\label{timing1}
&&\Delta\dot{p}
\simeq - {P_0\over T}\theta 
\ls \frac{}{} \cot\chi\lp \frac{}{} a_0\cos(2\pi t/1016)+ a_1\cos(2\pi t/508) 
\frac{}{}\rp \right.\no\\
&&\hspace{5cm}\left. 
 +\frac{\theta}{4}\lp\frac{}{} 2a_0a_1\cos(6\pi t/1016) + a_1^2
 \cos(2\pi t/254)
 \frac{}{}\rp\frac{}{}\rs,
\eea
where $t$ measured in day.  Here we ignore the $1000$ d and $500$ d
contributions
to the $\theta^2$ order.   
For $|a_0|\ll |a_1|$, the 250-component will be comparable to the
500-component if we have 
$(a_1\theta/4)\tan\chi>1 $, or $\tan\chi>4/(a_1\theta)$.   
For a small $\theta$ ($a_1\geq 1$), one finds that the magnetic dipole 
moment must be nearly 
orthogonal to the symmetry axis $\unit_3$.   Hence for $\chi>89^\circ$ and 
$\omega_p+\omega_d\simeq 2\pi/500$ d$^{-1}$, 
the most dominant terms are the second and forth harmonics, $500$ d and 
$250$ d, that is in good agreement with observed data.
Since the proposed inclination angle between star's spin axis and magnetic
field's symmetry axis is nearly right angle, $\chi>89^\circ$,
one may consider the rotation of dipole vector as a magnetic poles reversal.
We will get back to this point later.

It is interesting to note that equation (\ref{timing1}) includes $1000,
500, 333$, and $250$ d 
Fourier components. Further, by considering $n\geq 2$ terms in equation 
(\ref{spin-var-mag1}), 
one can find the 
higher harmonics in $\Delta\dot p$.   
These terms were missing in the Link \& Epstein's model. 
In table 1 we compare the time-varying magnetic field model with
one suggested by
Link \& Epstein (\cite{LE01}),
and the observations made by Stairs et al. (\cite{SLS00}).


\section{Discussion}\label{discuss}
In this paper, motivated by the effective core-crust decoupling
during the precessional motion (Link \& Cutler \cite{LC01}),
we considered the case of time-varying 
magnetic field of the star.   Then we endeavored
to find out under what condition a {\it time-varying} magnetic
field is able to provide a consistent explanation for 
the reported PSR B1828-11 timing analysis by Stairs et al. (\cite{SLS00}).  
We studied the free precession of spin axis of PSR B1828-11 under
the magnetic radiation torque caused by an inclined
time-varying magnetic dipole moment vector $\m(t)$,
with {\it constant} inclination angle $\chi$.  In general, 
we assumed that the dipole field vector changes with time relative
to the star's body axes.
Then we expanded its
components in terms of the Fourier expansion as 
$m_1=\m\cdot\unit_{1}=m_0\sin\chi\sum_{n=0} a_n\cos(n\omega_d+n\beta_d)$,
$m_2=\m\cdot\unit_{2}=m_0\sin\chi\sum_{n=0} b_n\sin(n\omega_d+n\beta_d)$, and  
$m_3=\m\cdot\unit_{3}=m_0\sin\chi\sum_{n=0} c_n\cos(n\omega_d+n\beta_d)$, 
where $a_n$, $b_n$, and $c_n$ were determined by fitting the data.  
Finally, we showed that if 
$\omega_d/\omega_p \simeq P_{\rm pre}/508-1$, one may consistently
explain the
whole observed spectrum of the Fourier power analysis of PSR B1828-11.
We find an acceptable fit to the data with a precession 
period of $1015$ d,
a wobble angle 
$\theta=3^\circ.2$, and the inclination angle $\chi=89^\circ$ 
between the magnetic dipole and star's symmetry axis.  Note that the chosen 
Fourier expansion coefficients are $a_0=.01$, $-b_1=a_1=1=c_0$, and 
$a_n=b_0=b_n=c_{n-1}=0$ for $n\geq 2$.

The time-varying magnetic field model can also explain the observed
timing data for PSR B1642-03.  The analysis of timing data of
PSR B1642-03, collected over a
span of $30$ years, exhibit strong Fourier power at periods $\simeq 5000, 
2500$, and $1250$ days (Shabanova et al. \cite{SLU01}).  The suggested
wobble and magnetic
field inclination angles are $\theta\simeq 0^\circ.8$ and $\chi\simeq 
60^\circ$, respectively.  
Similar to PSR B1828-11, the spectra of PSR B1642-03 show wide spectral 
features at periods $\simeq 2500$ d and $1250$ d.  The pulse shape 
variations were not detected, probably, due to their small amplitudes. 
Furthermore, it is interesting to note that both PSR B1828-11 and
PSR B1642-03
exhibit 
features around the sixth harmonic, $6\omega_p$, ie. $167$ d and $667$ d, 
respectively.  As seen from equation (\ref{spin-var-mag1}), by including
the $n= 2$ 
term in the expansion, one can reasonably get these harmonics in
$\Delta\dot p$.
These terms were missing in previous studies.

The necessary equality relation between $P_d$ and
$P_{\rm pre}$ requires the magnetic poles
at the surface of the star move with relative velocity
$v_{\rm rev}\sim 2\pi R/P_{\rm pre}\sim
7\times 10^{-2}$ cm s$^{-1}$ respect to the body axes.
On the other hand, the magnetic poles
reverse every $P_{\rm pre}/2\sim
500$ d.  Because of the very high electrical conductivity of the solid
crust ($\sigma\sim 10^{26}$ s$^{-1}$), this result would be hardly
acceptable.\footnote{
Though such a short magnetic cycle has not been observed in neutron stars yet, 
the early observations of A-type stars ($\alpha$-variables),
with kilogauss magnetic
field strength, showed large amplitude, nearly symmetric magnetic reversals 
in periods ranging from $4$ to $9$ days, close to the periods of the stars 
(Babcock \cite{Bab58}).
Several recent observations from the young rapidly rotating stars 
confirmed the existence of the solar-type magnetic cycle with $P_0/P_{\rm 
cyc}\simeq 10^{-4}$  
(Brandenburg et al. \cite{Bra98}, Kitchatinov et al. \cite{Kit00}).  
Of course these stars presumably have active convection
zone, for the case PSR B1828-11 with proposed period for
the magnetic cycle, we have $P_0/P_{\rm cyc}\simeq 10^{-9}$,
which is smaller
by $5$ orders of magnitude relative to one obtained for the young
rotating stars.    This may agree with the fact that in
neutron stars the convective fluid motions are hardly excited.}
Variation of the magnetic dipole vector with time in a neutron star (with
solid crust and no active convective zone)
may be understood through
the so-called {\it neutron star crustal tectonics} scenario which has been
proposed originally to explain magnetic dipole evolution and resulting
observable features in millisecond pulsars, low-mass X-ray binaries and
radio pulsars (Ruderman \cite{Rud91ab}).
The solid crustal lattice of neutron star is subject to various strong
stresses.  The pinned superfluid neutron star vortex lines exert a strong
force on the crustal lattice of nuclei which pin them
(Anderson \& Itoh \cite{And75}, Ruderman \cite{Rud76},
Alpar et al. \cite{Alp84}).
Further, the evolving core magnetic flux
tubes which pass through the crust, pull it strongly at the
base of the crust Srinivasan et al. (\cite{Sri90}).   In the
rapidly rotating weakly magnetized
neutron stars such as millisecond pulsars and low-mass X-ray binaries, the
lattice stresses from pinned vortices are dominant, while in the older
pulsars such as radio pulsars with
strong magnetic fields, the magnetic stresses from core flux-tube
displacement may become important.  For PSR B1828-11 and PSR B1642-03 with
magnetic field strength $B\sim 10^{12}$ G
and effectively unpinned superfluid vortices (Link \& Cutler \cite{LC01}),
the latter case
is more appropriate.  The quantized magnetic flux tubes in a core's type II
superconducting proton sea terminate at the base of the crust.  These flux
tubes move in response to changes in the positions of neutron star's core
superfluid.  
If the crust were to remain immobile the shear stress, $S(B)$, on the base
of the
highly conducting crust (from core magnetic flux tube motion) could
grow to reach
\st\be\label{cond}
S(B) \sim \frac{BB_c}{8\pi}\sim \lp\frac{B}{3\times 10^{12}\;{\rm G}}\rp
10^{26}\;\; {\rm dyn\; cm}^{-2}.
\ee
Here $B_c\geq 10^{15}$ G is the average magnetic field in each core
flux-tube, and $B$ is the average magnetic field through the crust.
The maximum
stress that crust could bear before breaking depends on the lattice shear
modulus and is calculated by Ruderman (\cite{Rud91ab}) as 
$S_{\rm max} \leq 10^{26}$ dyn cm$^{-2}$
(for most stars $S_{\rm max}\sim 10^{23}-10^{24}$ dyn cm$^{-2}$).   If
$S(B)>S_{\rm max}$, neutron stars with strongly magnetized cores would
break their crust continually as they rotate. For PSR B1828-11
with $B\sim 5\times 10^{12}$ G, we have $S(B)>S_{\rm max}$.
Therefore, one would expect a continuous crust breaking and
crustal plate motion in this pulsar.\footnote{It is interesting to
note that according to the neutron star crustal tectonics
scenario the magnetic
fields in spinning down neutron stars move to achieve a right angle
configuration relative to the star's spin axis (Ruderman \cite{Rud91ab}).
This is in
agreement with our analysis for PSR B1128-11 as we found
$\chi\geq 89^\circ$.}  In addition, the precessing
crust drags the core magnetic flux tubes which are frozen into the core's
fluid.  This would increase core flux-tube displacements and then
{\it increase}
the characteristic velocity of conducting platelets.   A typical
characteristic velocity of a flux-tube array in the stellar core
layer just below the core-crust interface is given by
Ruderman et al. (\cite{Rud98})
as $v_c \sim (\omega_0/10\;{\rm Hz})(10^{12}\;{\rm G}/B)^{-1} 10^{-7}$
cm s$^{-1}$.
For PSR B1828-11, $v_c \sim 3\times 10^{-8}$ cm s$^{-1}$
which is much smaller than the proposed relative velocity for
magnetic poles by our calculations,
$v_{\rm rev}\sim 2\pi R/P_{\rm pre}\sim
7 \times 10^{-2}$ cm s$^{-1}$.
We note that in calculation of $v_c$ the effect of
precessional
motion of the crust was not considered.  By including the precessional
effects, one may expect the value of flux-tube velocity $v_c$ to increase 
significantly.

It is worth noting that the required torque variation may as well be that
due to the internal torque, by the partially pinned vortices during the
precessional motion of the crust. The internal
torques would arise from different coupled components of the star which
move with different velocities, eg. the mutual friction torque which
arise from different velocities of vortex lines and superfluid.  These
torques are able to sustain hydromagnetic shear flows and turbulences
in the core-crust boundary, excite the fluid convection motions, and
cause magnetic field variations (Malkus \cite{Mal63}, \cite{Mal68}).
Further, they affect the motion of the neutron star crust, for example,
by tilting away its angular velocity vector from alignment with star 
principle axis (Sedrakian et al. \cite{Sed99}).  To find a clear
picture of dynamics of
magnetic field in a precessing neutron star, one has to consider the
effect of the internal torques.    This is currently under investigation
(Rezania \cite{Rez02}).
 
Finally, in this paper we showed that a time-varying magnetic field model
is able to explain consistently the timing analysis of both
PSR B1828-11 and PSR B1642-03, if the field's symmetry axis rotates
with a rate nearly equal to their precession rates, relative to the star's
body axes.
Unfortunately, at this stage, the large speed of the
magnetic poles at the surface of the star required by this model is
difficult to accept.
Further studies on the evolution and dynamics of magnetic fields
in precessing stars (especially the plate tectonics model) seem necessary.   
These will be left for future investigations.

\begin{acknowledgements}
I would like to thank S. M. Morsink , B. Link, and S. Sengupta for
their careful reading
the manuscript and useful discussions.  It is a pleasure to thank
M. Jahan-Miri
for stimulating discussions during the course of this work and draw
my attention to the plane tectonics model.
The author is also grateful to I. Stairs for useful discussions and
providing
the timing data for PSR B1828-11. I would like to thank Roy Maartens 
for continuing encouragement.  This research was supported by the Natural 
Sciences and Engineering Research Council of Canada.
\end{acknowledgements}

\newpage
\noindent {\bf Table Captions:}\\
\begin{itemize}
\item{Table 1:} 
In this table we compare the
time-varying magnetic field model for PSR B1828-11
with one suggested by 
Link \& Epstein (\cite{LE01}) and the observed data reported by Stairs et 
al. (\cite{SLS00}).  
$P_0$, $P_{\rm pre}$, 
$\epsilon=P_0/P_{\rm pre}$, $\theta$, and $\chi$ are the star's period, 
precession period, 
star's oblateness, wobble angle, and field's inclination angle, 
respectively. No 
observational information is available for the beam pattern.
\end{itemize}

\begin{center}
Table 1\vspace{.52cm}\\
\begin{tabular}{lllll}\hline\hline
Data/Model&$P_{\rm pre}$ (d)&$\epsilon$ &$\theta$&$\chi$\\\hline
Data  & $\simeq 1000,\, 500,\, 250,\, 167$ &  $4.7 \times 10^{-9}$
& $\simeq 3^\circ$ & $> 89^\circ$ \\                  
Time-varying mag. model & $\simeq 1000^{\rm a}$, all harmonics &  $4.7 
\times 10^{-9}$
& $\simeq 3^\circ$ & $> 89^\circ$\\
Link \& Epstein's model & $\simeq 500^{\rm a}$, $250^{\rm b}$ &  $9 
\times 10^{-9}$
& $\simeq 3^\circ$ & $> 89^\circ$\\ \hline 
& & & &  \\
$^{\rm a}$Fundamental&  & & & \\
$^{\rm b}$Harmonic & & & & 
\end{tabular}
\end{center}


\begin{thebibliography}{}

\bibitem[1984]{Alp84}
        Alpar, M. A., Andersson, P. W., Pines, D., \& Shaham, J. 1984, ApJ,  
               282, 791

\bibitem[1975]{And75}
        Andersson, P. W., \& Itoh, N., 1975, Nature, 256, 25 

\bibitem[1958]{Bab58}
        Babcock, W. H. 1958, ApJ, 128, 228

\bibitem[1998]{Bra98}
        Brandenburg, A., Saar, S. H., \& Turpin, C. R. 1998, ApJ, 498, L51


\bibitem[2001]{JA01}
        Jones, D. I., \& Andersson, N. 2001, MNRAS, 324, 811

\bibitem[2000]{Kit00}
        Kitchatinov,L.L., Jardine,M., \& Donati,J.-F. 2000,    
       MNRAS, 318, 1171

\bibitem[2002]{LC01}
        Link, B., \& Cutler, C. 2002, MNRAS, in press, 
   preprint (astro-ph/0108281)

\bibitem[2001]{LE01}
        Link, B., \& Epstein, R. I. 2001, ApJ, 556, 392

\bibitem[1963]{Mal63}
        Malkus, W. V. R. 1963, Journal Geophysical Research, 68, 2871

\bibitem[1968]{Mal68}
        Malkus, W. V. R. 1968, Science, 160, No. 3825, 259

\bibitem[2002]{Rez02}
        Rezania, V.  2002, work in progress

\bibitem[1976]{Rud76}
        Ruderman, M. 1984, ApJ, 203, 213 

\bibitem[1991a,b]{Rud91ab}
        Ruderman, M. 1991a, ApJ, 366, 213; 1991b, ApJ, 382, 587 

\bibitem[1998]{Rud98}
        Ruderman, M., Zhu, T., \& Chen, K. 1998, ApJ, 492, 267 

\bibitem[1999]{Sed99}
        Sedrakian, A., Wasserman, I., \& Cordes, J. M. 1999, ApJ, 524, 341

\bibitem[2001]{SLU01}
     Shabanova, T. V., Lyne, A. G., \& Urama, J. O. 2001, ApJ, 552, 321

\bibitem[1977]{Sha77}
        Shaham, J. 1977, ApJ, 214, 251

\bibitem[1990]{Sri90}
         Srinivasan, G., Bhattacharya, D., Muslimov, A., \& Tsygan, A. 1990, 
              Current Sci., 59, 31

\bibitem[2000]{SLS00}
     Stairs, I. H., Lyne, A. G., \& Shemar, S. L. 2000, Nature, 406, 484

\end{thebibliography}
\end{document}